\documentclass[prl,floatfix,twocolumn,showpacs]{revtex4-1}

\usepackage{graphicx}
\usepackage{subfigure}
\usepackage{amsmath}
\usepackage{amssymb}
\usepackage{latexsym}
\usepackage{multirow}
\usepackage[colorlinks=true]{hyperref}
\usepackage{blindtext}
\usepackage{pst-node,graphicx}

\begin{document}

\title{Mechanical instabilities of biological tubes.}

\author{Edouard Hannezo$^1$, Jacques Prost$^{1,2}$, Jean-Fran\c cois Joanny$^1$}
\affiliation{$^1$Physicochimie Curie (Institut Curie / CNRS-UMR168 /UPMC), Institut Curie, 
Centre de Recherche, 
26 rue d'Ulm 75248 Paris Cedex 05 France}
\affiliation{$^2$E.S.P.C.I, 10 rue Vauquelin, 75231 Paris Cedex 05, France}

\date{\today}

\begin{abstract}

We study theoretically the shapes of biological tubes affected by various pathologies. When epithelial cells grow at an uncontrolled rate, the negative tension produced by their division provokes a buckling instability. Several shapes are investigated : varicose, enlarged, sinusoidal or sausage-like, all of which are found in pathologies of tracheal, renal tubes or arteries. The final shape depends crucially on the mechanical parameters of the tissues : Young modulus, wall-to-lumen ratio, homeostatic pressure. We argue that since tissues must be in quasistatic mechanical equilibrium, abnormal shapes convey information as to what causes the pathology. We calculate a phase diagram of tubular instabilities which could be a helpful guide for investigating the underlying genetic regulation.
\end{abstract}
\pacs{xx}

\maketitle

Tubular structures are found ubiquitously in living organisms, from worms to humans and
are fundamental structures of many organs: they convey liquids, gases or cells 
throughout the body. Typical examples are arteries, intestinal tubes 
or renal excretory canals. The disruption of these tubes is at the origin of 
many pathologies, and although a considerable amount of research has focused
on the molecular mechanisms underlying each disease and each type of tube 
\cite{1,2,3,4,5}, little attention has been given to the biomechanical aspects 
of tube stability. Here, we  present a general framework to describe the instabilities 
of cellular tubes; we also discuss the possibility to infer
the underlying causes of pathologies by studying the shapes of the abnormally deformed tubes.
Indeed, tube instabilities are observed in many different organs but always seem to fall into a 
limited number of physiological categories : enlarged, sinusoidal, or 
varicose tubes\cite{2}\cite{6}. From a physicist's point of view, these shapes are reminiscent 
of classical fluid instabilities. Nevertheless, a key difference from passive systems is that pathological instabilities result from the internal activity of the tube.  Cell division, or active fluid pumping, are often the motor of the instability. Thus, a 
mechanical description of the tube shapes could bear a lot of information 
on the underlying mechanisms of tube formation and instability.

In the past years, progress has been made in the mechanical  
description of tissues as dividing elastic media\cite{7,8,9}. 
Stresses inside the growing tissues are essential to determine the final architecture 
of an organ and conversely, there are feedback mechanisms of form on growth. 
Here, we apply these models, which have been usually used for bulk tissues, 
to cylindrical geometries. We consider a wide variety of epithelial tubes that  all  
share key features. The tube is composed of a layer of dividing epithelial cells, which exert pressure on the surrounding medium, and of an elastic basement membrane 
which provides mechanical stability. A softer, visco-elastic 
connective tissue, which we consider infinite, is surrounding the tube.

This simple description suggests a competition between several forces. The fluid-tube interface has a tension and can undergo a Rayleigh-Plateau like instability
\cite{10}, reminiscent of a cylindrical jet breaking into drops. However, the tube is elastic, and can resist this deformation as discussed for passive lipid tubes \cite{11}.
Here we wish to add active effects: epithelial cells grow (sometimes at an uncontrolled rate), and the tube can buckle under the load induced by growth. This can be modeled by a uniform negative surface tension: cell divisions are driving an increase of the surface area of the tube. As noted in \cite{7}, sufficient mechanical feedbacks can explain why growth is spatially uniform. We will make that assumption here, given the regularity of the patterns we wish to explain. Tubes can also dilate because of an excess uniform fluid pressure.
\begin{figure}[!h]
\centering
	\includegraphics[width=8.5cm]{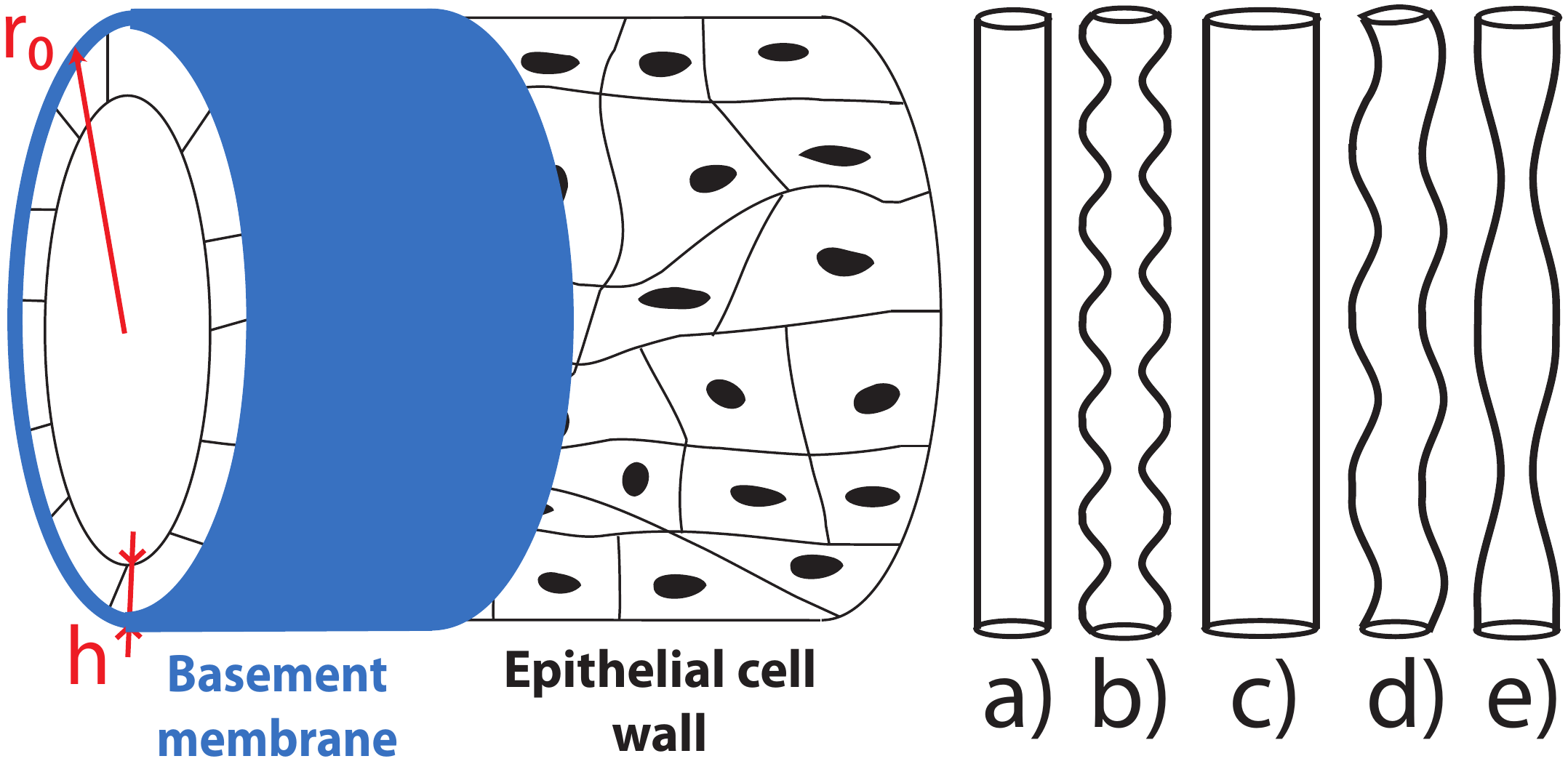}

	\caption{Left: Sketch of our model for biological tubes. A single layer of epithelial cells rests on an elastic membrane (blue), surrounded by soft connective tissue (not drawn). Right: Different tube configurations: a) reference, b) varicose/pearling, c) enlarged, d) sinusoidal e) sausage-like. b) and e) have the same symmetry, but correspond to different instabilities, as described in the text.}
	\label{figure1}
\end{figure}

We successively investigate the morphologies sketched on Fig.\ref{figure1}, in that order.
The Young modulus $E_t$ for biological tubes, such as arteries \cite{13}, is of the order of $10^4 - 10^6 Pa$, which is stiff compared to the surrounding tissues.
The reference state is a tube of infinite length and radius $r_0$. We first discuss a peristaltic perturbation (varicose, or pearling) which respects the cylindrical symmetry. The displacements can therefore be written $(u_r (z), u_z (z))$ where $z$ is the coordinate along the tube and $r$ 
the radial direction. 
The forces involved are equivalent to those derived from the following effective energy
\cite{12}:
\begin{equation}
\frac{{{\cal{E}}_{el}}}{2\pi r_0} = \frac{E_t h}{2} \int (e_z^2+e_{\theta}^2 + 2 \nu e_z e_{\theta}) dz  
+ \frac{K}{2} \int \left({\cal{C}} - \frac{1}{r_0}\right)^2 dz
\end{equation}
\normalsize
where $\nu$ is the Poisson ratio, and the strains are defined as
\begin{equation}
e_{\theta} = \frac{u_r}{r_0}+\frac{1}{2}(\frac{u_r}{r_0}^2) \, ;\, e_z = u_z'+\frac{1}{2} ( {u_r'}^2+{u'_z}^2)
\end{equation}
\normalsize
where the prime denotes a derivative with respect to $z$. The first term is the stretching elastic energy of the tube. We use here the classical 
Foppl-Von Karman approximation \cite{12}, neglecting $ {u'_z}^2$, but we keep the nonlinear term $(\frac{u_r}{r_0})^2$, if uniform dilations of the tube are non-negligible.
The second term is the layer bending energy where  $\cal{C}$ is the local curvature, and $K = \frac{E_t h^3}{12 (1-\nu^2)}$ the bending modulus. A key difference with Ref. \cite{50} is that we do not assume that the outer surface of the cylinder is fixed.
The case of  arteries is complicated by the presence of smooth muscles and anisotropy but as we wish to capture the essential physics, we restrain ourselves to this simple nonlinear and isotropic elastic theory. Surface and pressure forces, which drive the deformation, are derived from an effective energy in which the non-equilibrium aspect is hidden in the surface tension and excess pressure terms:  
\begin{multline}
\small
{\cal{E}}_a =  \int ( 2 \pi \gamma (r_0+u_r(z))\sqrt{1+u'_r(z)^2} \\- \pi
\Delta P (r_0+u_r(z))^2 ) dz
\end{multline}
\normalsize
where $\gamma$ is the effective surface tension, which contains a negative contribution due to cell division. As a result, $\gamma$ can be positive or negative. $\Delta P$ is an excess pressure inside the tube.
We first investigate a deformation of the tube to a radius $u_r (z) = r_0 [R_0 -1 + A\, cos(kz)]$ which includes both a uniform dilation to a radius $R_0 r_0$ and a pearling deformation of amplitude $A r_0\ll r_0$. All lengths have been normalized by $r_0$ (see SI for details).
We first discuss the effect of negative surface tension, assuming that there is a small excess 
pressure, $\Delta P r_0\ll |\gamma| (h/r_0)^2$. 
Compared to the classical Rayleigh-Plateau instability, the wavelength observed in vivo is small, close to the value of the tube radius \cite{2,6,14}. We propose that this pearling  instability is due to a buckling of the tube induced by the homeostatic pressure of the dividing epithelial cells. This hypothesis has biological grounds for hepatic arteries, since varicoses are associated with Fibromuscular Dysplasia (FMD)\cite{14, 15, 16}  or polycystic disorders in renal canals \cite{21}.

\begin{figure}[!h]
\centering
	\includegraphics[width=8cm]{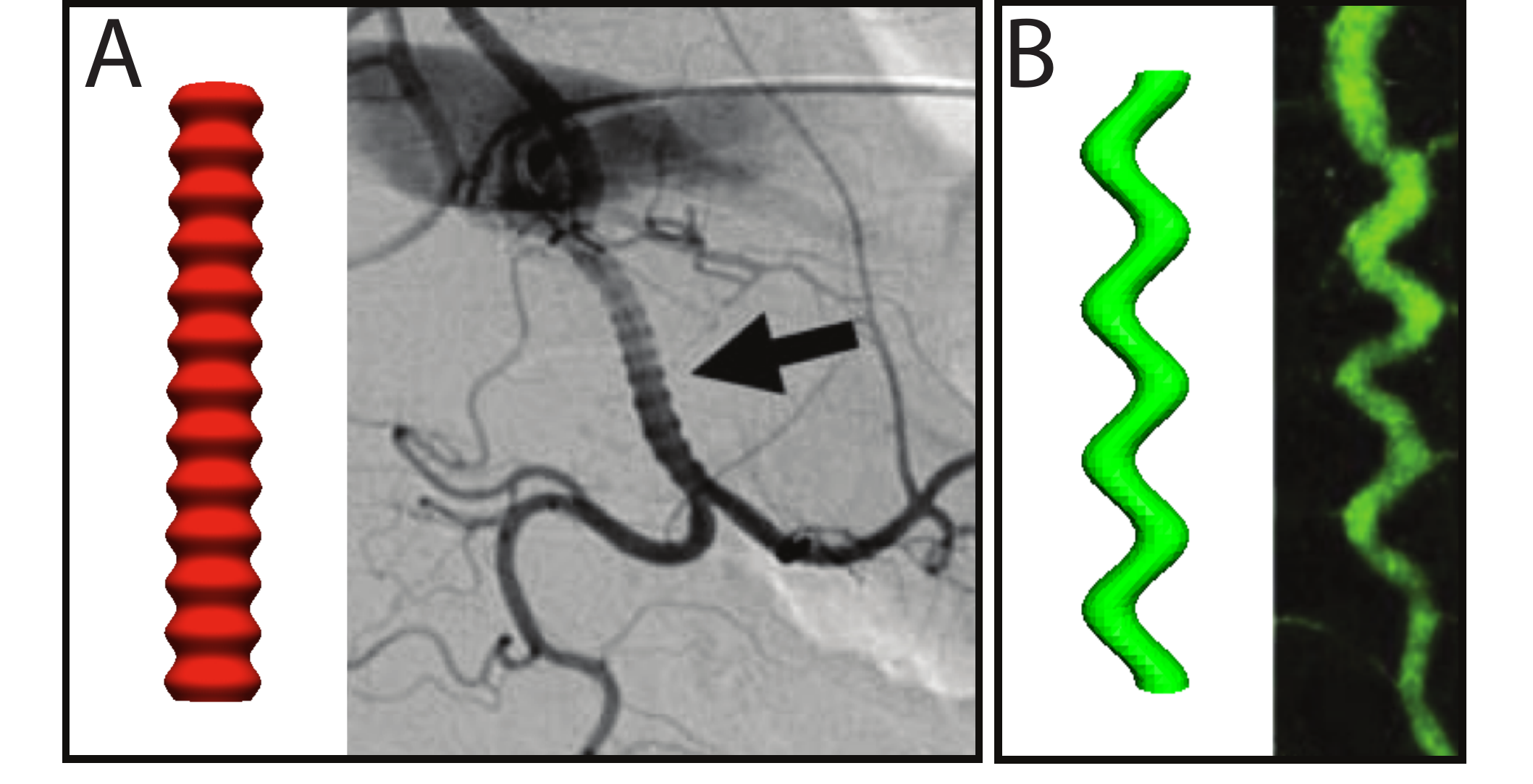}

	\caption{Comparison between the shapes obtained from buckling theory and in vivo data. A: a common hepatic artery displaying varicosities \cite{14}. B:  a drosophila tracheal tube displaying sinusoidal mutation \cite{45}, $\lambda \approx 5 r_0$.}
	\label{figure2}
\end{figure}

In the following, we assume that the value of the  tension is such that $\frac{h^2}{r_0^2} \ll \frac{|\gamma|}{E_t h}\ll 1$, which seems to be the case for reasonable values of the parameters : tubes we will study have $\frac{h^2}{r_0^2} \approx 0.01 $, and as we will discuss below, $\frac{|\gamma|}{E_t h} \approx 0.1$ . The minimization of the energy with respect to $R_0$ leads to $R_0 = 1 + \frac{|\gamma|}{E_t h}\simeq 1$. 
In this limit, there is a bifurcation from tubular to pearling states for 
\begin{equation}
\small
\label{pearling}
|\gamma| > \gamma_p = E_t h \frac{h}{\sqrt{3}r_0}
\end{equation}
\normalsize
at a wavelength $\lambda_p = 2\pi r_0 (\frac{1}{1-\nu^2}\frac{h}{\sqrt{12} r_0})^{\frac{1}{2}}$. The  finite elasticity of the surrounding medium does not change these equations as long as its modulus $Es$ is such that $ E_s r_0 \ll E_t h$, which is the case here. The Poisson modulus $\nu$ only has a weak effect. We have performed a numerical minimization and Fig.\ref{figure2} shows a plot of the resulting shape of a tube that has undergone a pearling instability. 
For large arteries such as the common hepatic artery, the wall to lumen ratio is 
approximately $\frac{h}{r_0}=0.1$, so that  $\lambda_p \approx 1.1 r_0$. Strikingly,  clinical studies \cite{15} confirm that the size of the pattern is proportional to the radius of the artery, with the exact same coefficient as above, in contradiction to analyses based on the  Rayleigh-Plateau instability \cite{17}. 

Away from the bifurcation point, the wavelength varies only slowly. This variation is 
compatible with the biological variability, and explains why the same range of wavelengths 
is always observed.
Since the wavelength is of the order of the radius, we need not consider instabilities in the section plane of the tube, studied in Ref. \cite{40}. For much thinner tubes, such an analysis would be required. 
The negative tension exerted by the cells is linked to the homeostatic pressure, 
$\gamma \approx - P_h h $ \cite{18}. Pressure measurements for tumors 
or dysplasia often give values in the range of $P_h = 10^4 Pa$ \cite{19}. Taking $E_t=10^5 Pa$, the critical tension $\gamma_p \approx 7.10^3 h$ is perfectly compatible with a  buckling induced by the dividing cells.

Polycystic diseases are more complex. The tubular canals of the kidney loose 
their stability, giving rise to a variety of patterns : string of small pearls, large convolutions, or huge spheric cysts that can invade the entire organ \cite{6} \cite{20}. 
Three mains factors \cite{21} have been identified : increase in basement membrane
compliance, uncontrolled division of the epithelial cells forming the tube wall, 
and disorders in ions pumps, causing excess water to be pumped in the canals. 
In our formalism, this relates respectively to a decrease in $E_t$, an increase in $P_h$,
and an increase in $\Delta P$. The excretory canal of C. Elegans is often used as 
a model organism to study these diseases. Although the tube is different in nature \cite{22} 
(a single, elongated cell invaginates to form a tube which spans its entire length), 
it can be described by similar equations : there is a bending energy associated 
to membrane deformation, a surrounding elastic medium prevents stretching, the 
tube can grow because of excessive lipid transport, analogous to negative surface 
tension, and there can exist hydrostatic pressure differences.
Thus, in order to incorporate the effect of pressure differences, we now consider the opposite limit of enlarged tubes, where the pressure dominates over tension effects, $\Delta P r_0 \gg |\gamma| (h/r_0)^2$ . 
The equilibrium radius is such that the elastic deformation equilibrates the excess pressure.
\begin{equation}
\small
R_0= \left(1+2 \frac{\Delta P}{E_t}\frac{r_0}{h}\right)^{1/2}
\end{equation}
\normalsize
The physical picture of pearling is not modified greatly by considering the effect of an 
excess pressure. The threshold tension remains of the same order, and 
the characteristic undulation wavelength is slightly reduced : $\lambda_p = \lambda_p (\Delta P = 0)(1+ 5 \Delta P r_0 / 3E_t h)^{-1/4}$.

We then turn to sinusoidal oscillations of the entire tube. The classical Euler buckling instability 
of an elastic column under compression occurs at a wavelength which is the system 
size, in the absence of a surrounding elastic medium. However, in an elastic medium,
buckling occurs at a finite wavelength. We have shown recently that this instability 
can be invoked to describe the morphology of the intestine \cite{23}.

We call $E_t$ and $E_s$ respectively the Young moduli of the tube and surrounding 
elastic medium, keeping in mind that $E_t >> E_s$. We denote the deformation of 
the tube by $w(z,t)=B\, cos(kz)$. The curvature modulus of a hollow elastic tube 
is $K= \pi E_t r_0^3 h$, the force exerted by the growing tube is 
$f=2\pi \gamma r_0$, and the elastic force exerted by the surrounding medium is 
proportional to the deformation with a coefficient $\alpha$.
The threshold force of the sinusoidal instability is $f= 2 \sqrt{K \alpha}$, thus the critical tension for buckling is
\begin{equation}
\small
\label{sinuous}
|\gamma| >\gamma_{s} \approx 0.8 \sqrt{r_0 h} \sqrt{E_s E_t}
\end{equation}
\normalsize
The wavelength of the instability is $\lambda_{s} = 2 \pi r_0 ( \frac{h}{2r_0} \frac{E_t}{E_s})^{\frac{1}{4}}$.
A rough estimate leads to  $\lambda_c \approx 5-10 \,r_0 $, in agreement with 
observations of degenerate excretory canals in C. Elegans \cite{6}, or of the tracheal system
in drosophila (\cite{2} and Fig.\ref{figure2}).

We now compare sinusoidal deformations, which occur for cell tensions larger than $\gamma_{s}$ given by Eq.(\ref{sinuous}), with pearling deformations with a threshold
$\gamma_{p}$ given by Eq.(\ref{pearling}). The preferred morphology essentially depends on the value of the parameter $r =\frac{E_s r_0^3}{E_t h^3}$. If the substrate is very soft or the tube very hard, the sinuous morphology is expected, since in that case, it requires a smaller tension. Conversely, if $\frac{h}{r_0}$ is very small, the pearling instability is expected. Physically, 
this is due to the fact that a larger tube is harder to bend than a thicker tube : 
its curvature modulus varies as the cube of the radius and as the first power of thickness.
For a typical tube, $\frac{h}{r_0}=0.1$, and $E_t=10^5 Pa$, the critical substrate rigidity separating both morphologies is $E_s = 10^2 Pa$.
This analysis is important since a loss of tubular rigidity is one of the causes of cystic 
diseases. The situation is 
much more complex if the substrate is modeled as viscoelastic. Over long timescales, 
sinuous deformations are always expected, but can be kinetically constrained. 
A detailed discussion goes  beyond the scope of this paper\cite{30}. 

In the region where both instabilities can be observed, it is necessary to compare the energies of the two configurations. We plot in thick green the transition line between the two instabilities, in the vicinity of the critical point.

\begin{figure}[!h]
\centering
	\includegraphics[width=9cm]{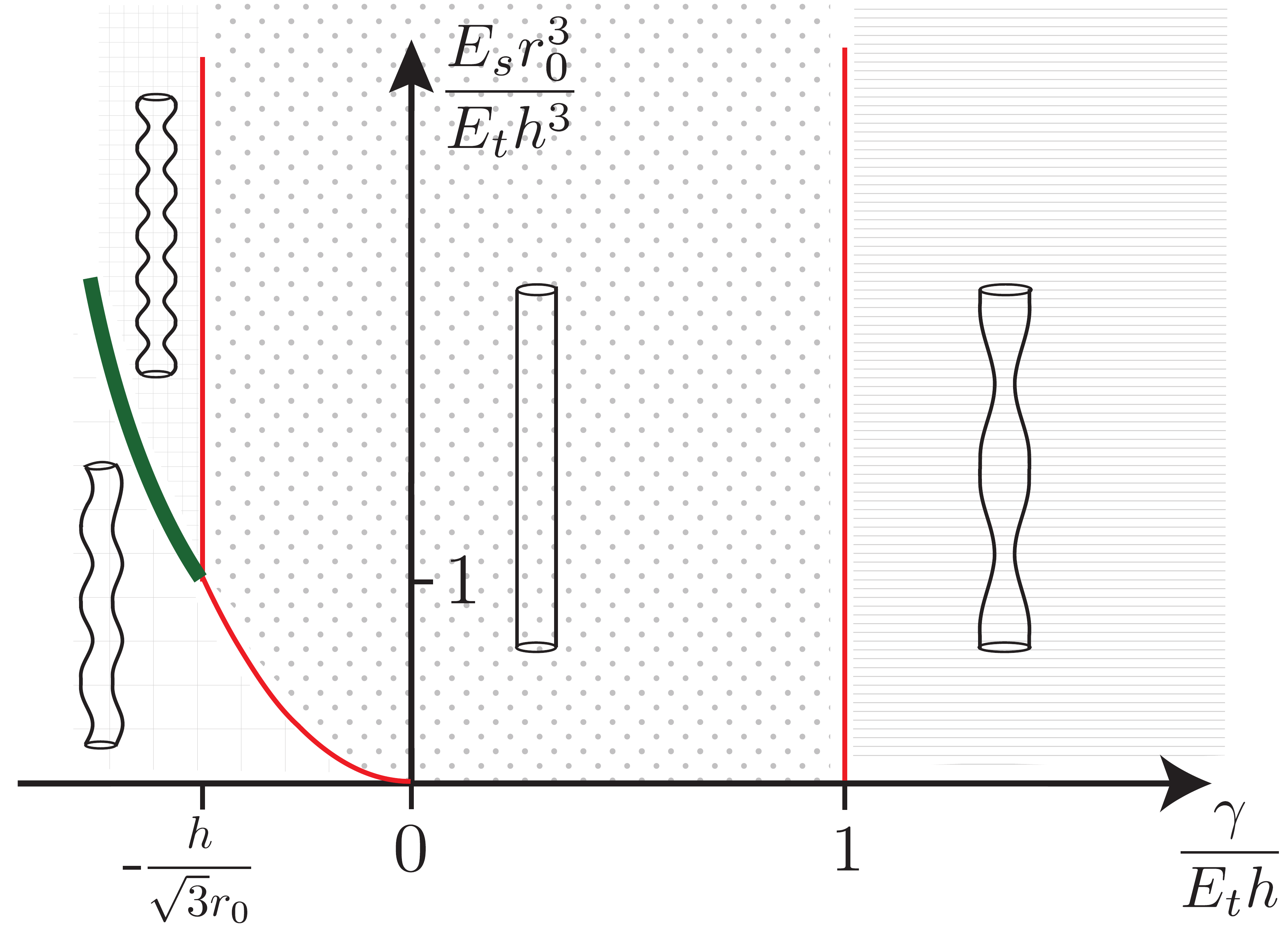}

	\caption{Stability diagram of all possible configurations of a tube, depending 
	on the relevant parameters. The vertical axis is the ratio of the moduli of the 
	surrounding tissue and of the tube. The horizontal axis is the ratio of the tension,
	negative or positive, and the elasticity of the tube. Lines in thin red represent the transition to an instability. The thick green line separates stability domains of the varicose and sinuous patterns. }
	\label{figure3}
\end{figure}
The results are displayed on a stability diagram in Fig.\ref{figure3}.
The drosophila tracheal system would be a good candidate for putting this theory to the test. 
It has been widely studied as a model of tube size regulation and the deletion of several  
genes has been identified to cause pathologies. Specifically, mutants lacking the Sinuous, 
or Varicose genes respectively display sinuous \cite{24}, and pearling state 
morphologies \cite{25}. Although we do not have a good understanding of their roles, 
it is known that they are required  in cell-cell junctions (associated to the rigidity of the tube)
and that they have a role in controlling growth. We now show how our phase diagram could guide further experiments. The modulus of the normal tracheal wall was inferred from experiments to be around $20 kPa$\cite{42}, and $\frac{h}{r_0} \approx 0.2$. Although tracheal connective tissue has not been probed, usual orders of magnitude are $100 Pa$. In normal organisms, the relevant ratio therefore is $\frac{E_s r_0^3}{E_t h^3} = 0.6 < 1$, so sinusoidal instabilities are expected. Nevertheless, disturbing junctions lowers the rigidity of the tube, and a three-fold decrease in $E_t$ would cause the pearling instability to be favored. We therefore suggest that more emphasis
should be put on measuring each gene influence on the mechanical properties of cellular tubes.

A last situation to address is a tube in the absence of any dysplasia of the cell wall. The surface tension is then positive. In this regime, we expect either a stable tube, or a classical Rayleigh-Plateau instability of an elastic cylinder. This is slightly different from the situation encountered in membrane tubes \cite{11}, since the fluid cylinder is surrounded by an elastic shell.

We study a perturbation $u_r(z)=r_0 [(R_0-1) + A\, cos(kz)]$. The situation that we wish to discuss does not involve any cell growth, and the instability thus occurs on a time scale small enough volume conservation is a reasonable assumption. This imposes that  $R_0^2-1= - \frac{A^2}{2}$. 
Minimizing the energy with respect to $A$ (SI) yields a finite value of $A$ only for a tension larger than the critical value $ \gamma_r = E_t h \frac{1-\nu^2}{1-q^2}$.
Below this critical value of the tension, the cylinder is stable. The instability first occurs for the longest wavelength mode $q \rightarrow 0$. Determining the exact most unstable wavelength requires a non-linear analysis, which is beyond the scope of this paper.

Nevertheless, if the size of the perturbation approaches the radius of the tube, disconnected pearls are obtained and the most unstable wavelength can be found by studying the dynamics of the perturbation $u_z$. The continuity equation and a Poiseuille approximation for the flow lead at lowest order to: $\frac{\partial u_z}{\partial t} = r_0^4/ (16 \eta)  \partial_z^2 P $ where $\eta$ the fluid viscosity. 
The hydrostatic pressure $P$ is calculated from the force-balance equation on the membrane $ 
P \propto \left( \gamma A (-q^2+1) - E_t h (1-\nu^2) A \right) cos(qz)$.
The most unstable wavenumber is then $q_r^2= \frac{\gamma - E_t h (1-\nu^2)}{2 \gamma}$.
This is an explanation similar to \cite{17} for the sausage-like pattern observed on some arteries.
Indeed it has been shown \cite{26} that a very soft elastic cylinder can undergo a classical Rayleigh-Plateau instability. Testing the case of hollow elastic tubes in the same fashion could be an experimental confirmation to our theory.
As expected, if $\gamma/ E_th \gg 1$, the most unstable wavevector converges
to the wavevector of the classical Rayleigh-Plateau instability $q_{RP} = \frac{1}{\sqrt{2}}$ \cite{10}.

The main result of this paper is  the stability diagram, sketched on Fig.\ref{figure3} 
presenting the stabilities of  most morphologies found in cellular tubes, depending 
of their mechanical properties. Buckling theory is in good quantitative agreement with the in vivo data. We have here deliberately used simple models for 
the mechanics of the tubes, to emphasize the generality of our results, and the relevance 
of a few dimensionless ratios to predict the observed instabilities. Most importantly, 
we want to stress the fact that tissues' shapes could turn out to be a meaningful criterion to detect the causes of specific pathologies. For example, in the case of excretory kidney canals, dysplasia leads to  a different instability than fluid accumulation. More work is clearly needed in each specific case, but we do not expect the basic physics presented in this paper to be modified.


\begin{thebibliography}{12}
\bibitem{1}
M. Affolter \textit{et al.}, Dev Cell, Vol. 4, 11Ð18, (2003)

\bibitem{2}
G. J. Beitel \textit{et al.}, Development 127, 3271-3282 (2000)

\bibitem{3}
M.A. Baer \textit{et al.},  Curr Top Dev Biol, Vol. 89 (2009)

\bibitem{4}
D. J. Andrew, A. J. Ewald, Dev Biol 341,34Ð55,  (2010) 

\bibitem{5}
B. Lubarsky, M. A. Krasnow, Cell, Vol. 112, 19-28 (2003)

\bibitem{6}
M. Buechner \textit{et al.} Dev Biol 214, 227-241 (1999) 

\bibitem{7}
B. Shraiman, P Natl Acad Sci USA, 102:3318-3323 (2005)

\bibitem{8}
T. Mammoto \textit{et al.}, Development 137:1407-1420 (2010)

\bibitem{9}
J. Ranft \textit{et al.}, P Natl Acad Sci USA, 107, 20863 (2010)

\bibitem{10}
Lord Rayleigh, Philos. Mag. 34, 145 (1892).

\bibitem{11}
R Bar-Ziv and E. Moses, Phys. Rev. Lett., 73, 10 (1994)

\bibitem{13}
S. Laurent \textit{et al.}, Arteriosclar Thromb Vasc Bio, 14:1223-131 (1994)

\bibitem{12}
S. Timoshenko, J.M. Gere, Theory of Elastic Stability, second ed. MacGraw-Hill, New York (1961).

\bibitem{50}
P. Ciarletta \textit{et al}, J. Mech. Phys. Solids, 60 525-537 (2012)

\bibitem{14}
B. Peynircioglu and B. E. Cil, Cardiovasc Intervent Radiol, 31:S38-S40 (2008)

\bibitem{15}
P. F. J. New, Am Roentgen Ray Soc, 97-1, 488-499 (1966)

\bibitem{16}
P.-F. Plouin et al, Orphanet Journal of Rare Diseases, 2:28 (2007) 

\bibitem{21}
L. W. Welling, Pathogenesis of cysts and cystic kidneys. In ÒThe Cystic KidneyÓ (K. D. Gardner and J. Bernstein, Eds.), 99-116. Kluwer, Amsterdam, Netherlands (1990)

\bibitem{17}
P. Alstrom et al, Phys. Rev. Lett, 82, 9 (1999)

\bibitem{40}
J. Yin \textit{et al.}, P Natl Acad Sci USA, 105,19132 (2008)

\bibitem{18}
M. Basan \textit{et al.}, HFSP Journal, 3(4):265-272 (2009)

\bibitem{19}
G. Helmlinger \textit{et al.}, Nat Med 3, 177- 82 (2007)

\bibitem{45}
P. Laprise \textit{et al.}, Curr Biol 12; 20(1): 55 (2010)

\bibitem{20}
A. P. Evan, J. A. McAteer, Cyst cells and cyst walls. In ÒThe Cystic KidneyÓ (K. D. Gardner and J. Bernstein, Eds.), pp. 21Ð42. Kluwer, Amsterdam, Netherlands  (1990)

\bibitem{22}
F. K. Nelson et al, J. Ultrastruct. Res. 82, 156Ð171 (1983) 

\bibitem{23}
E. Hannezo et al, Phys. Rev. Lett. 107, 078104 (2011)

\bibitem{30}
R. Huang, J. Mech. Phys. Solids, 53, 63-89 (2005) 

\bibitem{24}
V. M. Wu et al, J Cell Biol, Vol 164-2 (2004)

\bibitem{25}
V. M. Wu et al, Development 134, 999-1009 (2007)

\bibitem{42}
A.M. Cheshire et al, Dev Dyn 237:1874-2888 (2008)

\bibitem{26}
 S. Mora et al, Phys. Rev. Lett. 105, 214301 (2011)

\end{thebibliography}
\end{document}